# UML Model for Compressed Message Exchange

Ali B. Dauda[1], Baba S. Ahmed[2], Abubakar A. Idris[2], Audu M. Mabu[4], Ilyas I. Iliyas[5]

[1]University of Maiduguri, Nigeria
[2]Mai Idris Alooma Polytechnic, Nigeria
[3]Federal College of Agric Produce Technology, Kano, Nigeria
[4]Yobe State University, Nigeria
[5]Nigerian Defence Academy, Nigeria

ali.dauda@unimaid.edu.ng, fusam20@gmail.com, auwalabubakaridris@gmail.com, iliyasibrahimiliyas@gmail.com,

*Abstract*— Client-server is a model in distributed computing for exchanging information among machines. Web services is one of the technologies use for data exchange via network. The Web services is a collection of technologies that client-server model use also to exchange information. The Web services uses XML as the message wrapper to exchange information but the XML is always verbose and hence incurs latency in the communication. This paper used UML to model the Web services exchange for client-server applications. UML is a visual language for analysing the requirement of a software development. UML notations were used to offer the interactions between the objects and the elements of the Web services and how the information is compressed for the exchange. The analysis and the design of the requirements will be useful to Web services and cloud computing researchers and developers.

## I. INTRODUCTION

The client-server is a communication model for server-to-client information exchange. The model provides information or services to a client in a distributed systems framework via Internet [1]. Client-server model is widely used as model for information exchange in distributed system. The model comprises of protocols and technologies for information transfer. In distributed computing, services are utilized to operate on information. Services are functions or operations that are accessible via network to be reuse by applications or users. One of the technologies for client-server information exchange are the Web services [2].

Web services are a collection of protocols comprising of Simple Objects Access Protocol (SOAP), Web Service Description Language (WSDL), Universal Description, Discovery and Integration (UDDI) and the Extensive Markup Language (XML). Messages are formatted and tagged by the XML and use the SOAP as a protocol to transport the XML message over the Hypertext Transport Protocol (HTTP) [3]. Services are provided by application services referred to as Web services [2]. The client requests the services and the server produce the services based on the client request.

Several researches were conducted on information exchange using Web services [3],[4],[5]. In Web services, the client and the server request and provide the services respectively. The communication demands implementation of codes to facilitate the exchange process. Due to complex protocols and the request operations, analysis and design of the necessary requirement of the web services will aid in simplifying the coding of the functionalities. UML as the standard approach in describing the needed requirement will help to address both physical and logical elements in developing the Web services. Some studies have been conducted to simplify the process of software development using UML [6].

UML provides a visual language for the description of the artefacts of a software system. UML is a modelling language suitable for high-level modelling in which constructions can be mapped into code with the help of Model transformation tools [7]. UML is a visual language equipped with a rich set of diagrams [8]. Each kind of UML diagram provides a different perspective of the system to be developed. For instance, use case diagrams are suitable for requirement capture, state diagrams describe system states and transitions to be triggered for state changes and sequence diagrams are intended to be used for object-interaction design: message exchange and execution threads.

This paper presents the modelling of Web services compressed data messaging for a client-server interaction. The modelling is based on UML notations and diagrams. The diagrams used include: Use case, activity diagram, sequence diagram, class diagram and component diagram. The use case, activity diagram and sequence diagram were used for the analysis, while the class diagram and the component diagram were utilized in the design. The model represents the exchange of SOAP Web services message from server (the provider) to the client (the consumer).

The rest of the paper is organized as follows: Section 2 presents the review of related work on Client-server, Web services and UML. Section 3 presents the modelling of Web Services requirement analysis and design while Section 4 concludes the paper by presenting the conclusion and future work.

## II. RELATED WORK

Yu and Chen (2003) [3] defined Web services as a collection of protocols comprised of Simple Objects Access Protocol (SOAP), Web Service Description Language (WSDL), Universal Description, Discovery and Integration (UDDI) and the Extensive Markup Language (XML). Messages are formatted and tagged by the XML and use the SOAP as a protocol to transport the XML message over the Hypertext Transport Protocol (HTTP). Web services provides an open standard protocol for interoperability among nodes [9]. SOAP as a major protocol in the Web services, provides inter-communication among applications using XML-based messaging over the HTTP [10].

Web Services transforms web applications to the advance level of functionality [11]. It provides a standard way to realize system integration effectively using network [12]. It provides machine-to-machine interoperability communication across the web using an XML-based standard to create and consume the services by the provider and the consumer [13],[12].

Pearce and Poulovassilis (2009) [14] developed a client-server architecture for student-teacher feedback mechanism. The authors used UML to focus on the core functionalities in the implementation of the system. The server side consisted of four classes that uses HTTP to communicate with the client side that consisted of four classes. The authors use a sequence diagram to demonstrate the arrangement of the objects involve in the client-server communication. The study outcome is a Conceptual Model of student-teacher feedback system to aid effective communication in a school system.

Data compression is the compacting of information into smaller representative without missing its quality [15]. The compressed data is represented in digital form to save space or transported over a network [16],[17]. There are many data compression techniques with the same primary aim of reducing the size of the original data, and reconvert back to the same original form whenever required.

According to James Rum et al (2009) [7], UML is a standard notation for modelling the documentation of system analysis. It provides an easy transition for programmers to translate the design into programming language. Deepa Raj (2012) [18] used the UML to model the compression of image by using the class diagram and the sequence diagram. The author appropriately provided a generic segmentation of compress/decompress process of an image file. The modelling simplified the image compression process with the goal of aiding researchers and programmers to translate the design of the image compression into a programming code.

Jesus M., Almendros J., and Luis I., (2009) [19] presented an object-oriented interaction of user interface and the database based on UML. The authors used state diagram and sequence diagram to provide the analysis and the design of the interface and the database using a Model View Control (MVC)approach. The outcome of the modelling is a simplification of processes using UML to develop an online application.

## III. WEB SERVICES MODELLING REQUIREMENT ANALYSIS AND DESIGN

The purpose of this study is to identify the requirements and to model the interaction of the Web services system. This Section provides requirement analysis and design of the Web services by providing the interaction among various objects and components that will constitutes the functionality of an executable client server application based on Web services paradigm. In this study, the system analysis and design use UML's use-case diagram, activity diagram and sequence diagram class diagram and component diagram to model the exchange process.

*1) Use Case Diagram*

The use case depicted the functional reality of how the solution will work between two or more applications or machines in the web services transaction. It represents the functional requirement of the Web services showing how graphically the services can be requested and be consumed. Figure 1 shows the transaction between service provider and the service requester in the Web services. The figure is a graphical generalization for modelling Web service with any transport protocol.

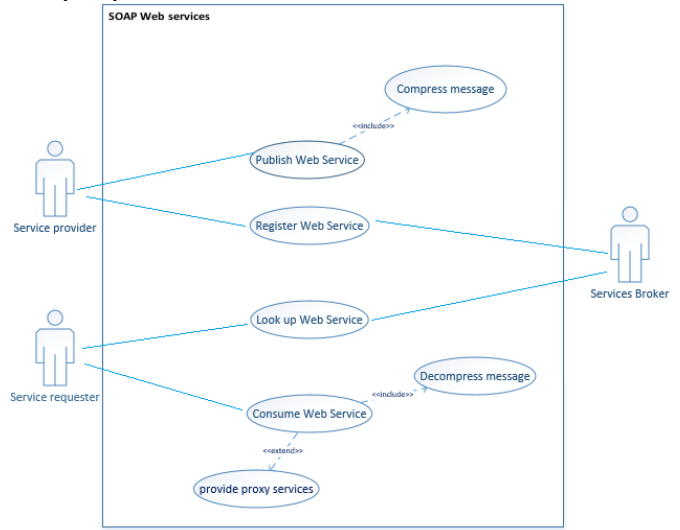

**Figure 1: Use case diagram for the SOAP performance Web services enhancement showing the detailed use cases and the actors of the Web services**

The use case diagram consists of three (3) actors and seven (7) use cases. The model uses publish-subscribe pattern to provide and consume the services. The *service provider* actor publishes the service by including the compress message use cases. The provider web service includes the *compress message* use case to compress the message if the message needs to be compressed. The *service requester* then subscribes via the *consumer web service* use case to consume the provided service by the *service provider* by including the *decompress message* use case to decompress the message provided by the *service provider*.

In the use case, the service requester request for service from the service provider. The service provider act by processing and compressing the message. The service provider publishes and register the service at the service broker. The service broker is the mechanism that resides all the published services and act as a mediator between the requester and the provider [20]. The requester interacts with the broker and check for services to be consumed [21]. The service requester uses a lookup service and utilize the provided services via a proxy link in the service broker. The service requester received and decompressed the message, and complete the transaction.

*2) Activity Diagram*

The activity diagram describes the dynamic workflow of the Web service. It shows the interaction between the provider and the consumer web services. This captured the initiation of the request to the stage when the service will be provided and consumed. Figure 2 illustrates the activities in the Web services indicating the activity flow in the system. The figure is a generalization diagram for both HTTP and JMS web services.

In this case, a request is performed by the requester web service and acknowledge by the provider web service. The provider web service processes the request by compressing the message and generating the corresponding WSDL file (22). This file contains a SOAP response wrapped in XML format specifying the location of the service and methods of the service, using a bounded protocol [23].

The requester web service receives the response and act by decompressing the message to its original form. The requester web service continually checks and processes the sent message until no further message is received from the provider web service.

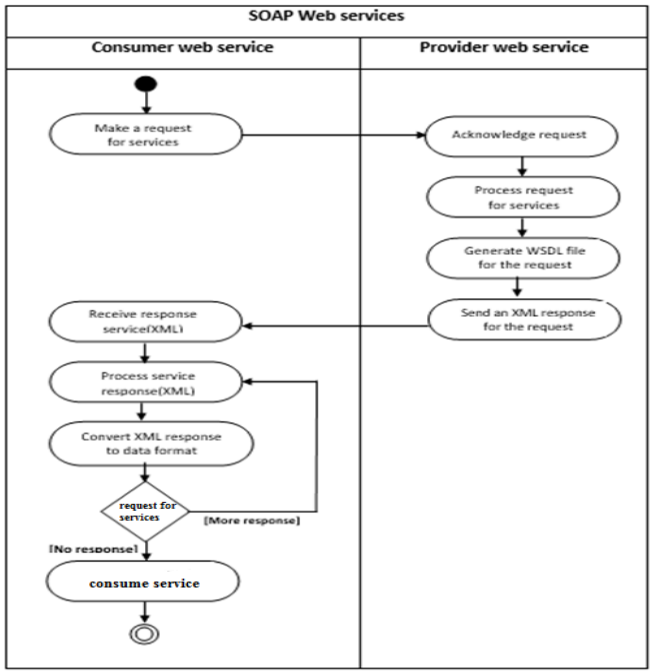

**Figure 2: Activity diagram for the SOAP performance Web services enhancement showing the dynamic flow of activities of the Web services.**

*3) Sequence Diagram*

A sequence diagram is utilized to model the dynamic behaviour of the objects in the solution of the Web services. Here, the sequence diagram models the sequential flow of objects participating in the Web services. It depicts how messages are exchanged in the Web services communication over time. Figure 3 shows the chronological procession of the objects from the requesting of the service to its consumption.

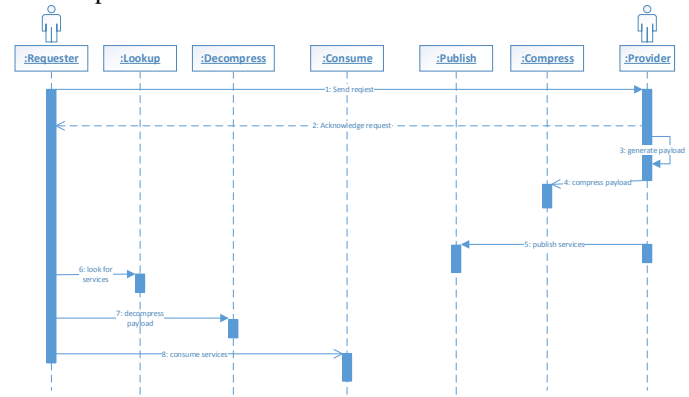

**Figure 3: Sequence diagram for the SOAP performance Web services enhancement showing the objects involved in the request-response implementation.**

The requester makes a request for service and the provider responds by acknowledging the request. The provider generates a message (payload) and compresses the message (if there is a need to be compress). The provider then publishes the compressed payload for consumption. The requester looks up for the services, if available the requester decompresses the payload and finally consumes the service.

Depicts the order in which the message is requested and consumed. In this sequence diagram, the transaction objects are; : Requester, :Lookup, :Decompress,: Consume, :Publish, :Compress and :Provider. The requester makes a request and the provider acknowledge the request, then process and compress the message. The provider publishes the message in the publishing service (Service broker). The requester uses a Lookup service to get the published service and then decompress the message and consume the services. As seen in the figure 4, the requester timeline stands active throughout the communication while the provider timeline stopped midway. Reason is that the requester makes the request and wait until the consummation of the services, while the provider will turn into inactive mode as long as the service is published. In this case, the provider generates message.

*4) Class Diagram*

The class diagram shows the stationary model view of the solution. It describes how the classes and their elements are arranged to work in the entire Web services experiment. As shown in Figure 4 the classes, methods, operations and constraints were related and modelled to capture and describe the responsibility of the system. The figure is the design consisting of controller class, web service provider class, web service client class and their corresponding compressor and decompressor classes. This model is designed to be transported via HTTP.

The class diagram is designed independent of programming language. It contained two endpoints: the provider web service and the consumer web service. The provider web services comprised of three classes namely, WSProvider, controller, and compressorClass. The controller class is the main method class. The class contained normalMsgHandler() method for handling the services for normal message transaction while the compressMsgHandler() method is responsible for handling compress message services.

The WSProvider class contain the all the necessary functions for establishing, connecting and providing the services to the consumer endpoint. In this case, message is generated by the provider and send to the appropriate message handler in the controller class. The time taken to generate, compress and send the message is recorded by the WSProvider timer () method.

The compressorClass is responsible for compressing the generated message. This class contained searchBuffer (), apppendBuffer (), readBuffer () and increasedBuffer () methods. The searchBuffer () concatenate the generated string to form a single message while the searchBuffer () is responsible for reading the string and encoding the symbols. But when the assigned buffer size is reached the increasedBuffer () method increase the length of the buffer to avoid buffer under run. This permits the readBuffer () method to accommodate more incoming symbols. The timer () records the activities at the process at the provider web service.

Conversely, the consumer web service comprised of four classes, namely, the WSConsumer and decompressorClass. The WSConsumer class has three methods: WSReceiverProxy (), WSProviderService (), onmessage () and the timer (). The WSReceiverProxy () invokes the provider web service and request for service. It acts as an intermediary between the web service provider and consumer. The WSProviderService () automate the connection to the Web service and consume the request. It contained several operations on how the provider web service is utilize and consumed.

The onMessage () receives the delivery of message send from the provider endpoint. The onMessage () method ensure that the received message is in the right format or otherwise cast the message to the appropriate format. The timer () records the activities at the process at the consumer web service.

In the decompressorClass is responsible for converting the received message back to original format. It contains three methods namely, compareBuffer (), apppendBuffer (), and increasedBuffer ().

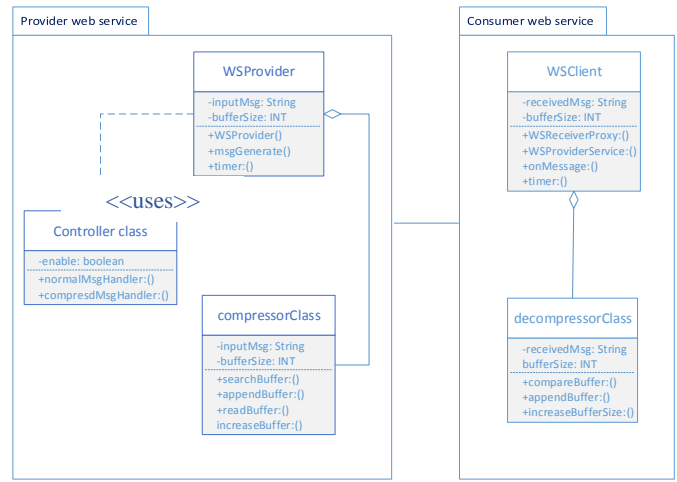

**Figure 4: Class diagram for the SOAP performance Web services enhancement showing the classes involved in the HTTP web services implementation**

*5) System Component Diagram*

The component diagram of the Web services shows the structural relationships between the components of the consumer and the provider web services. The components provide logical view of the artefacts that will lead to the implementation of the entire web services system. Figure 5 shows the component diagram for the Web services. The diagrams show how the entire provider/consumer services is modularized for effective control and ease of reuse. The figure depicts the component diagram for SOAP web service over HTTP transport. There are two subsystems in the figure: the provider web service and the consumer web service. The major artefacts in the provider web service subsystems are message provider, message compressor and service proxy. At the consumer web service subsystems, the artefacts are message decompressor, message consumer and invoker service. These two subsystems are connected to communicate via required/provided interface, basically as HTTP to exchange an XML-based message.

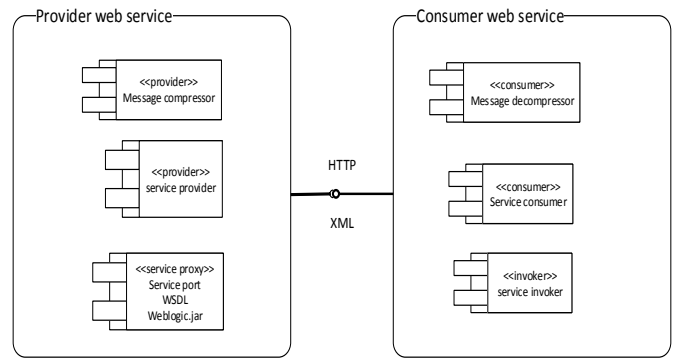

**Figure 5: Component diagram for the SOAP performance Web services enhancement showing the major components for achieving the functionalities of the implementation of HTTP web services**

## IV. CONCLUSION AND FUTURE WORK

The modelling of a software development process provides an easy transition from problem analysis to coding and implementation. The modelling using UML offers a description of the constituents involve in a particular software development life cycle and hence addressing the problem beforehand. Web services connect two or more applications to communicate and share services. The information exchanged by the web services can be affected by latency, thus there is need for improving the processing and transmission time. Data compression provides the compaction of information into a smaller version for effective communication. The UML notations were used to provide solution by modelling the interaction of web services elements that compress and exchange the information from server to the client. The UML model of this Web services will aid researchers and programmers to translate the design into programming code, which resultantly make the coding faster and provides efficient software. We anticipate to use a suitable data compression algorithm and SOAP Web services using Java to translate this UML models into a programming code that will effectively exchange the compressed message.